\PassOptionsToPackage{svgnames,dvipsnames,table}{xcolor}
\documentclass[sigconf]{acmart}
\renewcommand\footnotetextcopyrightpermission[1]{}
\AtBeginDocument{%
}
\usepackage[utf8]{inputenc}
\usepackage[T1]{fontenc}
\usepackage{textcomp}

\usepackage{amsmath,amsfonts,amssymb}
\usepackage{algorithm}
\usepackage{algorithmic}
\usepackage[inline]{enumitem}
\usepackage{graphicx}
\usepackage{booktabs}
\usepackage{multirow}
\usepackage{tcolorbox}
\usepackage[font=footnotesize]{caption}
\usepackage{float}
\usepackage{url}
\usepackage{balance}
\usepackage{pifont}
\usepackage{listings}
\DeclareCaptionFormat{listing}{\parbox[t]{\textwidth}{#1#2#3}}
\usepackage{etc}
\usepackage{tikz}
\usepackage{framed}
\usepackage{lmodern}
\usepackage{listings}
\usepackage{colortbl}
\usepackage{balance}

\newcounter{findingctr}
  
\definecolor{shadecolor}{rgb}{0.94,0.96,0.98}
\definecolor{shadecolorCode}{gray}{0.98}
\definecolor{listinggray}{gray}{0.98}
\definecolor{listings-comment}{RGB}{0,120,0}
\definecolor{listings-keyword}{RGB}{0,0,180}
\definecolor{listings-string}{RGB}{170,0,0}
\definecolor{errorred}{RGB}{180,30,40}
\definecolor{strong}{RGB}{255,220,150} 

\definecolor{codegray}{rgb}{0.5,0.5,0.5}

\newcommand*\circled[1]{\tikz[baseline=(char.base)]{
            \node[shape=circle,draw,fill=black,text=white,inner sep=2pt] (char) {#1};}}

\lstdefinestyle{github}{
    backgroundcolor=\color[HTML]{FFFFFF},
    basicstyle=\linespread{1.1}\footnotesize\ttfamily\color[HTML]{333333},
    commentstyle=\color[HTML]{EF6565},
    keywordstyle=\color[HTML]{3498DB}\bfseries,
    numberstyle=\tiny\color[HTML]{BDC3C7},
    stringstyle=\color[HTML]{446EBB},
    identifierstyle=\color[HTML]{3E3F3D},
    numbers=left,
    numbersep=10pt,
    frame=leftline,
    framerule=0.8pt,
    framesep=8pt,
    framextopmargin=6pt,
    framexbottommargin=6pt,
    xleftmargin=10pt,
    rulecolor=\color{gray!30},
    fillcolor=\color[HTML]{F8F9FA},
    tabsize=4,
    breaklines=true,
    captionpos=b,
    showspaces=false,
    showstringspaces=false
}

\newtcolorbox{rqbox}[1][]{
    colback=green!10,
    colframe=codegreen,
    arc=1mm,
    boxrule=0.5pt,
    coltitle=black,
    fonttitle=\bfseries,
    title=#1
}

\newtcolorbox{rqboxgray}[1][]{
    colback=codegray!10,
    colframe=codegray,
    arc=1mm,
    boxrule=0.5pt,
    coltitle=black,
    fonttitle=\bfseries,
    title=#1
}

\author{Sigma Jahan}
\affiliation{%
  \institution{Dalhousie University}
  \city{Halifax}
  \state{Nova Scotia}
  \country{Canada}
}
\email{sigma.jahan@dal.ca}

\author{Saurabh Singh Rajput}
\affiliation{%
  \institution{Dalhousie University}
  \city{Halifax}
  \state{Nova Scotia}
  \country{Canada}
}
\email{saurabh@dal.ca}

\author{Tushar Sharma}
\affiliation{%
  \institution{Dalhousie University}
  \city{Halifax}
  \state{Nova Scotia}
  \country{Canada}
}
\email{tushar@dal.ca}

\author{Mohammad Masudur Rahman}
\affiliation{%
  \institution{Dalhousie University}
  \city{Halifax}
  \state{Nova Scotia}
  \country{Canada}
}
\email{masud.rahman@dal.ca}

\renewcommand\footnotetextcopyrightpermission[1]{} 
\acmConference[]{}{}{}
\begin{document}
\title{Taxonomy of Faults in Attention-Based Neural Networks}

\begin{abstract}
Attention mechanisms are at the core of modern neural architectures, powering systems ranging from ChatGPT to autonomous vehicles and driving a major economic impact. However, high-profile failures, such as ChatGPT's nonsensical outputs or Google's suspension of Gemini's image generation due to attention weight errors, highlight a critical gap: existing deep learning fault taxonomies might not adequately capture the unique failures introduced by attention mechanisms. This gap leaves practitioners without actionable diagnostic guidance. To address this gap, we present the first comprehensive empirical study of faults in attention-based neural networks (ABNNs). Our work is based on a systematic analysis of 555 real-world faults collected from 96 projects across ten frameworks, including GitHub, Hugging Face, and Stack Overflow. Through our analysis, we develop a novel taxonomy comprising seven attention-specific fault categories, not captured by existing work. Our results show that over half of the ABNN faults arise from mechanisms unique to attention architectures. We further analyze the root causes and manifestations of these faults through various symptoms. Finally, by analyzing symptom–root cause associations, we identify four evidence-based diagnostic heuristics that explain 33.0\% of attention-specific faults, offering the first systematic diagnostic guidance for attention-based models.
\end{abstract}

\keywords{Attention Mechanisms, Fault Taxonomy, Transformer, Root Cause Analysis, Silent Failures}
\maketitle
\section{Introduction}
Deep Neural Networks (DNNs) have revolutionized many application domains, including medical imaging ~\cite{medicalimaging}, finance, autonomous vehicles ~\cite{selfdrivingcar}, and software engineering~\cite{softwaredevelopment}. Although early breakthroughs were led by feedforward, convolutional, and recurrent neural networks, recent progress has been dominated by networks adopting attention mechanisms, particularly the transformer family~\cite{vaswani2017attention} (\eg{} GPT-4~\cite{gpt4}, Gemini~\cite{gemini}, Copilot~\cite{copilot}). Analysts expect these models to add approximately $1.01$ trillion to the global economy by 2031, growing 26.6\% annually~\cite{statistaArtificialIntelligence}. However, this rapid progress has also led to critical reliability issues that hinder large-scale adoption. In February 2024, ChatGPT (\ie{} GPT 4) produced nonsensical outputs due to token mapping errors in its attention mechanism in a faulty inference kernel \cite{chatgptinsane, gptgibberish,openaigibberish}. That same month, Google paused Gemini's image generation feature, since its miscalculations in attention weights led to historically inaccurate outputs \eg{} generating images of WWII German soldiers as people of color~\cite{reutersGemini2024, blogGoogleGemini2024}.

Attention-based Neural Networks (ABNNs) include both attention-augmented architectures (\eg{} CNN or RNN with integrated attention modules) and pure self-attention architectures (\eg{} Transformers)~\cite{attentionPlease, vaswani2017attention}. Throughout this paper, we use ABNNs to refer to both categories, focusing on the faults arising from attention mechanisms in either context. The use of attention mechanisms that selectively focus on different parts of an input introduces structural instabilities and behavioural anomalies that the classic architectures (\eg{} CNN, RNN) do not face~\cite{vaswani2017attention}. For example, two of such major issues are: \emph{attention collapse}, where attention weights become abnormally concentrated on a small subset of tokens~\cite{attentioncollapse}, and \emph{attention drift}, where query-key alignments progressively deteriorate and distort contextual focus~\cite{du2024drift}.

Existing studies focus on deep learning faults across various classic DNN architectures such as MLP, CNN, RNN, exploring issues in model layers, tensor operations, and training processes~\cite{dlbugcharacterstics, faulttaxonomy}. Other work examines faults across deep learning training pipelines, including framework-related issues~\cite{silentbugs-framework, performance-bugs-framework, pytorchbugs}, deployment failures~\cite{deploymentbugs}, aging-related  bugs~\cite{aging-related-bugs}, optimization-related faults~\cite{MLoptimizationBug}, and compiler-related issues~\cite{faultsinCompiler, BugsDeepLearningCompilers, comilerbugsempirical}. These studies have systematically defined and categorized faults, and established root cause-to-symptom mappings, which benefited the subsequent fault diagnostic tools (\eg{} DeepDiagnosis~\cite{deepdiagnosis}, DeepFD~\cite{deepfd}, Neuralint~\cite{Neuralink}, DEFault~\cite{default}). Although they have investigated various DNN architectures and advanced the state of knowledge, attention-specific faults remain largely unexplored and less understood, indicating a significant gap in the literature. 

To address this gap, we present the first empirical study of software faults in attention-based neural networks. We consider attention-specific faults as those that arise uniquely from attention-related components, such as attention masks and QKV projections. We attempt to better understand the attention-specific faults, their root causes, and symptoms, and thus derive actionable insights for software and deep learning practitioners.
We first collect 555 real-world faults from 96 projects, covering ten DNN frameworks, across GitHub, Hugging Face, and Stack Overflow. Then, we analyze these faults using a mix of qualitative and quantitative methods and answer four research questions as follows.\\
\textbf{RQ1: What are the unique fault types introduced by attention mechanisms in neural networks, and how prevalent are they?}\\
We found that over half of all faults (52\%) in ABNNs belong to seven novel fault categories, each arising exclusively from attention mechanisms and not covered by existing DNN taxonomies~\cite{faulttaxonomy, dlbugcharacterstics}.\\
\textbf{RQ2: What are the main root causes of attention-specific faults?}\\
We identified 25 root causes of attention-specific faults across seven fault categories in ABNNs. These include reasons such as QKV dimension mismatches, dynamic mask generation faults, and kernel memory overflows, which frequently contribute to failures in attention mechanisms.\\
\textbf{RQ3: What are the common symptoms of attention-specific faults?}\\
We observed that attention-specific faults often cause \textit{silent} or \textit{latent} problems, such as output corruption or context leakage. These issues occur in ABNNs two times more than in classic DNNs and can spread without being detected, leading to incorrect or unreliable results.\\
\textbf{RQ4: Can we systematically map root causes to observable symptoms in ABNNs to support fault diagnosis?}\\
We found statistically significant links between specific root causes and their observable symptoms in ABNNs. Four evidence-based diagnostic heuristics derived from these associations explain 33.0\% of attention-specific faults.\\
In summary, our work makes the following contributions.

(a) We conduct the \textit{first} comprehensive empirical study that systematically characterizes attention-specific faults in ABNNs, and develop a novel taxonomy with seven unique fault categories and 25 root causes, based on an analysis of 555 real-world faults. 

(b) We advance the understanding of fault diagnosis in ABNNs by analyzing the associations between symptoms and root causes, and propose four evidence-based diagnostic heuristics to facilitate more systematic fault detection.
 
(c) We make public our dataset of attention-specific faults and provide a comprehensive replication package to enable reproducible research, available at~\cite{anonymous2025}.
        

\section{Motivating Example}
\label{sec:motivating_example}
Existing fault taxonomies might address attention-specific faults only at a coarse-grained level, providing little guidance for diagnosis or resolution. To demonstrate the gap, consider the real-world issue~\cite{hf-19045} shown in Listing~\ref{lst:attention_fault}.

\begin{lstlisting}[
    style=github,
    language=Python,
    numbers=left,
    caption={Example of an attention-specific fault},
    captionpos=t,
    label={lst:attention_fault}
]
# Fault-triggering config
config = BertConfig(is_decoder=True, position_embedding_type='relative_key', max_position_embeddings=512)
model = BertLMHeadModel(config)
model.config.use_cache = True
# Input & first pass (establish cache)
input_ids = torch.tensor([[101, 2054, 2003, 1996, 3291, 102]])
output1 = model(input_ids)
# Extended input (uses cache, triggers fault)
extended_ids = torch.tensor([[101, 2054, 2003, 1996, 3291, 2000, 102]])
output2 = model(extended_ids, past_key_values=output1.past_key_values) # Silent output corruption
\end{lstlisting}

The example shows a fault in the BERT model that uses relative positional encoding with caching enabled\\(\ie{} \texttt{position\_embedding\_type} set to `relative\_key', line 3; \\\texttt{use\_cache} set to True, line 6). The model first processes an input sequence to establish the cache (lines 8--9). Then it processes an extended input with the cached key-value pairs (lines 11--12). However, the model fails to update the relative positions for new tokens, causing the attention mechanism to compute incorrect attention weights. As a result, the output is silently corrupted, with no error or warning. 

When we try to classify this fault using existing taxonomies, we find several key limitations as follows.\\
(a) According to \citet{faulttaxonomy}'s DL fault taxonomy, the example falls under `Model' faults. However, this high-level categorization provides no hints on the interactions between positional encoding and caching, which are central to this failure.\\
\noindent
(b) According to \citet{dlbugcharacterstics}'s work, the example fault will be treated as a `logic error' in model implementation. While accurate, this classification lacks the granularity to differentiate among different types of attention-related state management failures and offers no guidance for targeted debugging.\\
\noindent
(c) Studies focusing on framework-related faults~\cite{silentbugs-framework, pytorchbugs} would correctly identify our example as a silent fault. However, these taxonomies only address faults within the framework code (\eg{} tensor operations, kernel faults) and do not address errors introduced at higher abstraction levels, such as model configuration or architectural composition.

As shown above, none of these existing taxonomies can adequately capture the critical interaction among positional encoding variants (relative vs. absolute), stateful caching mechanisms, and attention weight computation dependencies. As a result, current taxonomies offer only limited guidance for diagnosing or resolving such faults in attention-based neural networks.

\begin{figure*}[ht]
  \centering

  \includegraphics[width=0.80\textwidth]{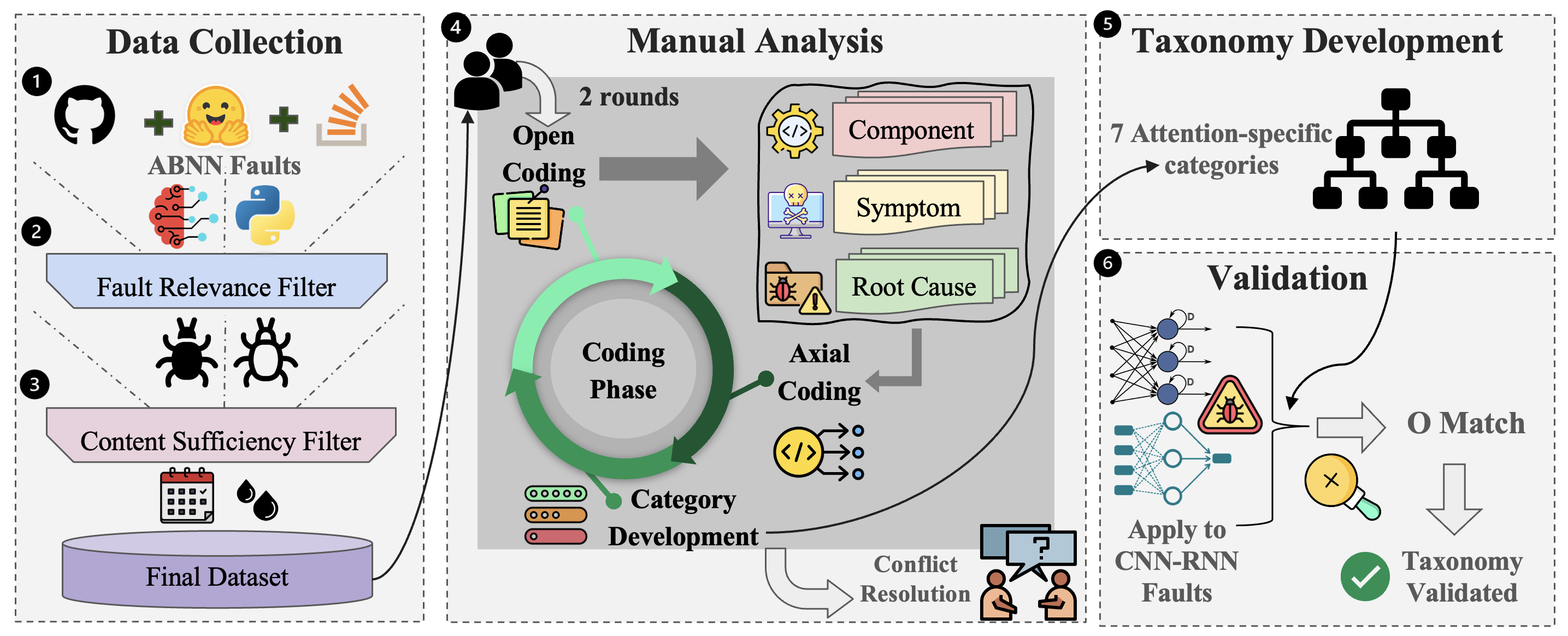}
  \Description{Schematic diagram showing the workflow for taxonomy development.}

  \caption{Schematic diagram of our workflow for taxonomy development}
   \vspace{-1.5em}
  \label{fig:dataprep}
\end{figure*}

\section{Methodolody}

We conduct an empirical study to better understand the faults, their symptoms, and root causes in attention-based neural networks (ABNNs). Fig.\ref{fig:dataprep} shows the schematic diagram of our conducted study. We discuss the major steps of our methodology as follows.

\subsection{Data Collection}
\label{subsection:data_collection}
\subsubsection{Initial Retrieval}
As shown in step \circled{1}, we collect bug reports, issue discussions, and programming Q\&A posts (\ie{} each referred to as an \emph{artifact}) that reference attention-based neural networks (ABNNs) from three sources widely used in DNN fault research~\cite{deepfd, deepdiagnosis, autotrainer}: GitHub Issues, Hugging Face forums, and Stack Overflow. We carefully maintain the following criteria in our data collection: 

(a) We target only Python-related bug reports and discussions due to Python's dominance in neural network development~\cite{octoverse2024}.

(b) We adopt a keyword-driven retrieval approach, following prior studies~\cite{faultsinCompiler, faultsinRL}. We select five attention-specific keywords (\eg{} \texttt{attention}, \texttt{attn}, \texttt{self-attention}, \texttt{multihead}, \texttt{transformer}) based on their prevalence in the attention mechanism literature~\cite{vaswani2017attention}.

(c) We restrict our retrieval to closed GitHub issues, ensuring that only developer-confirmed and resolved faults are included, following existing work~\cite{autonomousBug, faultsinRL, faultsinCompiler}.

(d) We limit Stack Overflow posts with accepted answers and scores $\geq$5, ensuring their quality and community validation as recommended by prior studies~\cite{dlbugcharacterstics, deeplocalize}.

(e) We include Hugging Face threads with at least two replies, as multi-reply discussions indicate meaningful engagement with the thread, following current practices~\cite{khan2024insights, wu2024characterizing}.

(f) We collect artifacts posted between January 2021 and December 2024 to capture recent faults and failures during the widespread adoption of transformer-based models. We identified 861 artifacts in the initial retrieval stage.
\subsubsection{Fault-Relevance Filtering}
Since the initially retrieved artifacts could be noisy, in step \circled{2}, two authors with more than four years of experience in DNNs and software engineering independently analyzed all artifacts using a predefined selection checklist (see replication package~\cite{anonymous2025}). For each artifact, we first excluded irrelevant ones, such as generic discussions, feature requests, tutorials, and theoretical posts. This initial filtering yielded 812 artifacts. We then carefully analyzed the remaining artifacts, retaining only those that explicitly reported faults in models with attention mechanisms. This included faults in attention-specific operations (\eg{} attention score calculation, mask handling, multi-head attention configuration) as well as supporting processes relevant to attention models (\eg{} embedding integration, hardware-optimized attention kernels). We achieved high inter-rater agreement for this inclusion step (Cohen’s $\kappa = 0.93$), resolving all disagreements through discussion. We also discarded artifacts describing faults in classic, non-ABNN architectures (\eg{} MLP, pure CNN, vanilla RNN). This filtering step reduced our dataset to 773 ABNN fault artifacts.

\subsubsection{Content Sufficiency Filtering}
Following the hybrid process established in prior work~\cite{verdi2020empirical, zhang2019empirical}, for step \circled{3}, we applied appropriate heuristics, such as automated detection of code blocks, error messages, or fault-related keywords, to filter out artifacts lacking technical detail. After this automated screening, two authors independently reviewed each remaining artifact using a predefined sufficiency checklist (see replication package~\cite{anonymous2025}). We considered an artifact sufficient if it included at least one of the following: an error message or stack trace, a code snippet, or a clear textual fault description related to attention mechanisms~\cite{faultsinRL, faulttaxonomy, dlbugcharacterstics}. We resolved any disagreements through discussion until reaching a consensus. We also removed duplicate faults reported across three platforms. This final filtering delivered 555 unique attention-specific fault instances from 96 Python projects across 10 deep learning frameworks. Please note that our dataset of 555 fault instances is substantial compared to that of similar studies. One of the seminal works on the taxonomy of DNN faults analyzed 375 fault instances~\cite{faulttaxonomy}.

Table~\ref{tab:data_stat} presents the diversity of our collected faults using a representative sample of 100 fault instances. We selected this sample using stratified random sampling~\cite{stratifiedRandomSampling} across four dimensions: framework, attention mechanism, model type, and development stage. We use this sample only to display dataset diversity due to the paper's space limitations, and we perform all subsequent analyses on the full set of 555 artifacts. The replication package~\cite{anonymous2025} contains complete diversity statistics for the full dataset.

\subsection{Manual Analysis}
\label{subsec:manual_analysis}
To systematically analyze attention-specific faults and answer our research questions, we employed an open coding approach as shown in Step \circled{4}, following established guidelines from software engineering research~\cite{braun2006thematic, cruzes2011recommended}. Two authors, each with more than four years of experience in deep learning and software engineering, annotated each fault instance across three dimensions: \textit{component} (the module or layer where the fault occurred), \textit{symptom} (the observable error manifestation), and \textit{root cause} (the underlying technical issue). We choose our codes based on the analysis of error messages, attached code snippets, and developer descriptions.

First, we generated an initial set of open codes for each dimension during a pilot review in which both authors jointly annotated 50 fault instances and achieved substantial agreement (Cohen’s $\kappa$ = 0.88). Following the pilot, both authors independently annotated the remaining fault instances, marking any cases as \texttt{Unknown} when sufficient information was not available. Throughout the entire coding process, we maintained a shared and versioned codebook for each dimension. When either annotator encountered a fault that did not fit the existing codes from the pilot review, they proposed a new code or modification. Both annotators discussed and reached consensus before incorporating any new or revised codes into the codebook. Whenever the codebook was updated, we systematically reviewed all previously annotated instances and jointly recoded any cases that could be more accurately described by the new or modified codes. This process ensured annotation consistency across the dataset and was repeated until thematic saturation was reached (\ie{} no new codes emerged in the final rounds of analysis)~\cite{braun2006thematic}.

Final inter-rater agreement scores for the three dimensions were Cohen’s $\kappa$ = 0.906 (component), 0.847 (symptom), and 0.928 (root cause). All remaining disagreements were resolved through discussion. The complete manual annotation process required approximately 555 person-hours for each annotator.

\begin{table*}[!t]
  \centering
  \caption{Dataset diversity summary for representative sample (N=100)}
  \vspace{-1em}
  \label{tab:data_stat}
    \resizebox{\textwidth}{!}{%
  \begin{tabular}{@{}l r l l r l l r l l r l@{}}
    \hline
    \multicolumn{3}{c}{\textbf{Framework}} &
    \multicolumn{3}{c}{\textbf{Attention}} &
    \multicolumn{3}{c}{\textbf{Model}} &
    \multicolumn{3}{c}{\textbf{Stage}} \\
    \cmidrule(lr){1-3} \cmidrule(lr){4-6} \cmidrule(lr){7-9} \cmidrule(l){10-12}
    Category & \# & Examples &
    Category & \# & Examples &
    Category & \# & Examples &
    Category & \# & Examples \\
    \hline
    PyTorch        & 42 & Llama, GPT        & Multi-Head       & 36 & BERT              & LLMs           & 30 & Llama-3          & Pre-training        & 26 & Foundation \\
    HuggingFace    & 33 & T5, BERT          & Self-Attention   & 26 & T5                & Diffusion      &  8 & Stable Diffusion & Fine-tuning         & 21 & Adaptation \\
    TensorFlow     & 11 & Keras             & Flash Attention  & 14 & LLMs              & BERT Variants  &  6 & DeBERTa          & Optimization        & 24 & TensorRT \\
    JAX            &  7 & Octo              & Cross-Attention  & 11 & Diffusion         & Custom Models  & 42 & Various          & Inference           & 17 & ONNX \\
    Others         &  7 & SHARK             & Sparse / HW      & 13 & Triton            & Multimodal     & 14 & CLIP             & Testing/Validation  & 12 & Benchmark \\
    \hline
  \end{tabular}
  }
\end{table*}

\subsection{Taxonomy Development}
To develop our attention-specific fault taxonomy, in Step \circled{5} we employed a bottom-up approach following previous work~\cite{faultsinRL, faulttaxonomy}. First, two authors independently grouped fault codes by the component dimension established during manual analysis (see Sec.~\ref{subsec:manual_analysis}), with each group forming a root category. For example,  all faults involving the attention masking component formed the \textit{Fault in Attention Masking} parent category. For each parent category, we then assigned the corresponding root causes as subcategories. For example, we assigned \textit{mask generation faults}, \textit{mask application faults}, and \textit{dynamic mask mismatches} as subcategories of the \textit{fault in attention masking} category. This initial independent classification yielded substantial agreement (Cohen’s $\kappa$ =0.91), after which all remaining differences were resolved through discussion. After reconciling these classifications, both the authors jointly reviewed the complete taxonomy hierarchy to clarify category names, refine definitions, merge any redundant subcategories, and ensure that parent–child relationships were logically organized and non-overlapping. Finally, we identified seven distinct fault categories representing attention-specific issues not covered by the existing taxonomies on DNN faults (see Fig.~\ref{fig:taxonomy}).

\subsection{Establishing Taxonomy Uniqueness}
To demonstrate the necessity and uniqueness of our taxonomy, we performed a comparative analysis using faults collected from classic DNN architectures. We collected 1,135 classical DNN faults by adopting the same retrieval criteria and mechanisms (see Sec.~\ref{subsection:data_collection}). Instead of attention-related keywords, we used a common set of keywords (\eg{} \texttt{conv}, \texttt{cnn}, \texttt{convolution}, \texttt{convs}, \texttt{convnet}, \texttt{conv2d}, \texttt{rnn}, \texttt{lstm}, \texttt{gru}, \texttt{recurrent}, \texttt{bilstm}, \texttt{brnn}) to target classic DNN architectures (\eg{} CNN, RNN). For simplicity, we did not include further variants and restricted our search to these common terms. Since manually analyzing 1,135 faults is extremely costly and 1,100+ person-hours have already been spent, we employed stratified random sampling to obtain a representative sample of 384 faults with a 95\% confidence level and 5\% margin of error. We sampled across three dimensions: (a) architecture type (CNN, RNN), (b) framework (\eg{} PyTorch, TensorFlow, Keras), and (c) platform source (GitHub, Stack Overflow, Hugging Face forums). Our sample includes 230 CNN-related faults, 115 RNN-related faults, and 39 hybrid architecture faults, distributed across 58 Python projects and nine frameworks covering the same timeline (\ie{} January 2021--December 2024). Two authors independently analyzed each fault using our attention-specific taxonomy and attempted to identify any instances of our seven categories. We achieved high agreement with Cohen's $\kappa{} = 0.92$. Notably, all disagreements across labelling tasks were resolved through structured discussions in physical meetings. The complete taxonomy and coding guidelines are available in our replication package~\cite{anonymous2025}. 

\section{Study Findings}
\begin{figure*}[ht]
  \centering
  \includegraphics[width=0.95\textwidth]{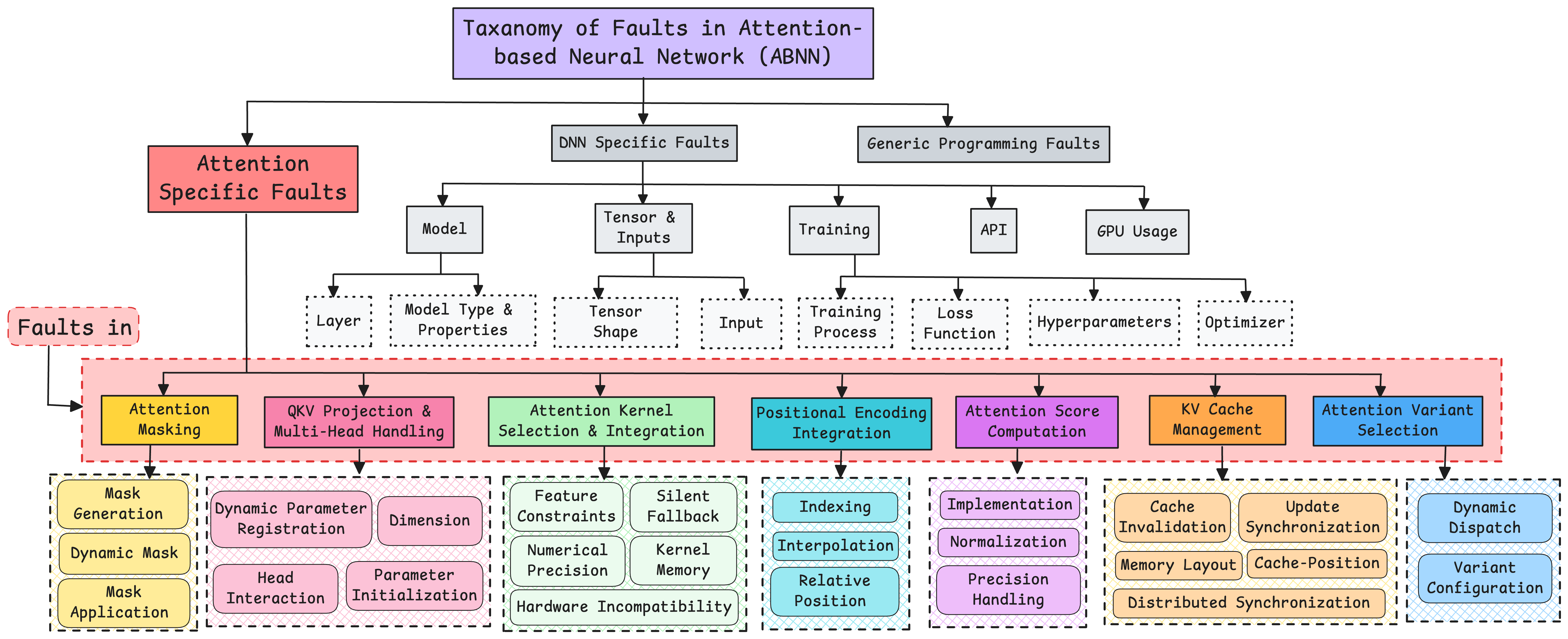}
  \Description{Taxonomy of faults in ABNN. Colored boxes indicate new fault categories, while light gray denotes existing ones.}
  \caption{Taxonomy of faults in ABNN. Colored boxes indicate new fault categories, while light gray denotes existing ones.}
  \label{fig:taxonomy}
\end{figure*}

\subsection{Answering RQ1: Fault Types} 
\label{sec:rq1_results}
To answer RQ1, we identify fault types \emph{unique} to attention mechanisms and analyze their \emph{prevalence} in ABNNs.

\subsubsection{Unique Fault Types}

\begin{figure}[h]
    \centering
    \includegraphics[width=\linewidth]{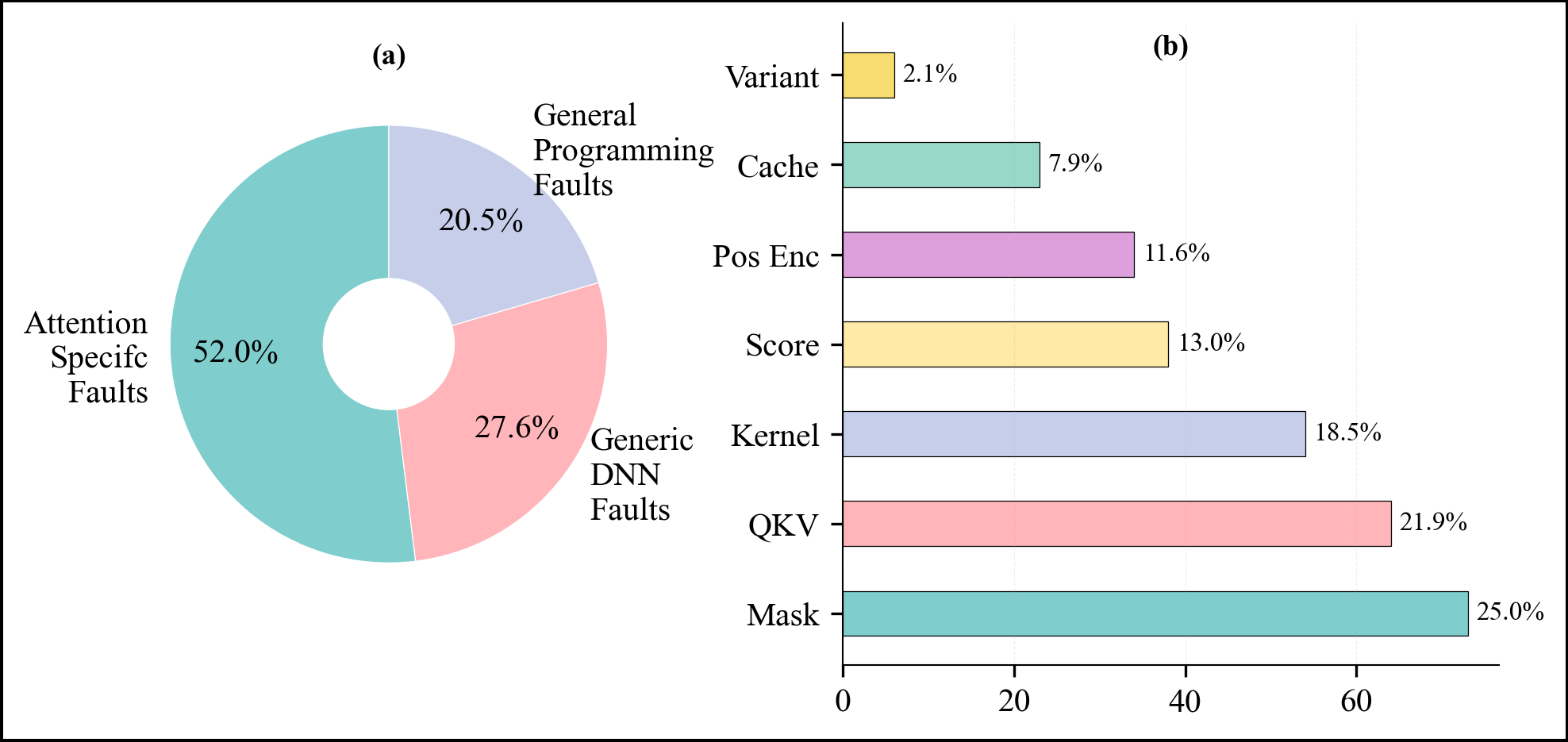}
    \caption{Distribution of (a) all faults in ABNN (b) attention-specific faults}
    \Description{Bar chart showing the distribution of all faults in ABNN and attention-specific faults.}
    \label{fig:abnn-fault-dist}
\end{figure}

We attempted to map 555 ABNN faults to existing fault taxonomies. We found that 27.6\% could be mapped to existing DL fault categories~\cite{faulttaxonomy}, 20.4\% were generic programming faults, and 52\% could not be characterized by any existing taxonomies. This motivated us to define new attention-specific categories, which account for 52\% of all observed faults (see Fig.\ref{fig:abnn-fault-dist}). 

To confirm the uniqueness of our seven attention-specific fault categories, we further analyzed a stratified random sample of 384 faults from classic DNN architectures (\eg{} CNN, RNN) (see \cite{anonymous2025} for details). Most of them mapped to classic DNN fault categories, such as tensor shape mismatches, incorrect hyperparameters, and training faults, but none matched any of the seven attention-specific categories, confirming their exclusivity.
We integrated our attention-specific taxonomy with established DNN taxonomies~\cite{faulttaxonomy}, following the approach of~\citet{faultsinRL}, to offer a comprehensive taxonomy for software practitioners (see Fig.~\ref{fig:taxonomy}). 

\subsubsection{Prevalent Fault Types}
We present the seven attention-specific fault categories (see Fig.~\ref{fig:abnn-fault-dist}) in descending order of prevalence within the 292 attention-related faults (52\% of all faults in ABNN).

Fault in \textbf{Attention Masking} is the most prevalent attention-specific category, accounting for 25\% of the 292 faults in our dataset. We identified these faults by analyzing explicit masking errors in bug reports and code patches. Based on our open-coding analysis, 51\% of masking faults resulted in high-severity failures, such as performance degradation or system crashes, and 34\% affected three or more downstream steps (\eg{} score computation, output logits, loss calculation).

Fault in \textbf{QKV Projection \& Multi-Head Handling} category ranks second, comprising 21.9\% of attention-specific faults. They arise when inputs are projected into separate query, key, and value spaces, and extra parameter matrices and shape constraints are introduced~\cite{vaswani2017attention, attentiondetails}. We found that 78\% of these faults can cause high-severity failures, indicating the vulnerability of the projection step to dimension mismatches and parameter initialization errors.

Fault in \textbf{Attention Kernel Selection \& Integration} (18.5\%) arises from incompatibilities between attention implementations and execution environments. While optimized kernels (\eg{} FlashAttention, SDPA, xFormers) improve performance and efficiency, they may also introduce hardware-specific and numerical errors that can cause subtle but critical failures, even when the attention algorithm is correct~\cite{memoryefficientflashattn, attentiondetails, pytorch2023sdpa}. As kernel implementations have become more specialized, setup and hardware requirements have grown more demanding~\cite{flashattention}. During manual analysis, we identified a steady increase in such faults, with their proportion rising from 18\% in 2021 to 42\% in 2024.

Fault in \textbf{Attention Score Computation} (13.0\%) arises at stages such as score calculation, softmax normalization, and multi-head aggregation~\cite{vaswani2017attention}. 
We found the three-stage process (\textit{compute}$\rightarrow$\textit{normalize}$\rightarrow$\textit{aggregate}) of computing attention score often creates tightly linked failure points, where faults frequently propagate across stages. To examine this propagation, we analyzed bug-fixing commits using an automated script based on existing work~\cite{diff, mondal2017bug}. The script mapped code changes to the respective stages and marked faults as propagated if the fixes spanned multiple stages. For example, in GitHub issue~\#31468, a misplaced dropout operation after the softmax normalization step corrupted the aggregation of the attention head outputs, showing how propagation can complicate debugging. 

Fault in \textbf{Positional Encoding Integration} makes up 11.6\% of attention-specific failures from our dataset. Although embedding layers often hold less than 1\% of a model's parameters~\cite{kaplan2020scaling, vaswani2017attention}, we found 78\% of these faults to be highly severe. They occur when positional embeddings are calculated for an incorrect or outdated position of a token, silently corrupting the model outputs. Even a single index drift can negatively affect all subsequent tokens~\cite{attentioncollapsecompress, attentiondetails}.

Fault in \textbf{KV Cache Management} accounts for 7.9\% of attention-specific failures and stems from the stateful execution of attention mechanisms during model inference. Unlike stateless feedforward models, attention-based models keep track of intermediate states across tokens as each generated token depends on previous ones, creating unique vulnerabilities~\cite{vaswani2017attention, attentiondetails}. We found that 65\% of the faults in \emph{KV Cache Management} category manifest during inference, where tokens are generated step by step, unlike training, where all tokens of a sequence are processed in parallel.

Fault in \textbf{Attention Variant Selection} is less frequent (2.1\%), but often associated with high severity. Unlike kernel faults, which arise from static incompatibilities between an attention kernel and the hardware or data types~\cite{memoryefficientflashattn, attentionengine}, these faults stem from dynamic, runtime choices of attention kernel implementations~\cite{attentionSurvey}. For example, transformer libraries may select one of FlashAttention, SDPA, and xFormers based on the input sequence length or hardware. If the selected variant differs from the configuration used for tuning, performance or accuracy can drop silently, violating the \textit{fail-fast} principle~\cite{shore2004fail}.

\begin{rqboxgray}
\textbf{Summary of RQ1:} More than half (52\%) of all faults in ABNNs are attention-specific and are not captured by existing deep learning fault taxonomies. We identified seven new fault types unique to attention architectures, with Attention Masking being the most common (25\% of cases) and Attention Variant Selection the least common (2.1\%).
\end{rqboxgray}

\subsection{Answering RQ2: Root Causes}
\label{sec:rq2_results}
RQ2 investigates the reasons (\ie{} root causes) behind attention-specific faults. To answer RQ2, we analyzed fault instances across all categories identified in RQ1, using bug fixes to determine their root causes.

\indent\textbf{Root causes behind fault in Attention Masking.}
Attention masks serve as critical gatekeepers for information flow, but our analysis reveals three main root causes for faults: (a) errors in how the mask is created (mask generation faults), (b) errors in how the mask is applied (mask application faults), and (c) failures in updating or synchronizing masks during sequential operations (dynamic mask mismatches).

The most common issues are \textit{mask generation faults}, which manifest through shape mismatches, missing parameters, faulty initialization, and incorrect static mask computation. We found that, in 16.5\% of attention models in our dataset, these issues produced masks that did not properly restrict information flow (\ie{} information leakage). Even when attention masks are correctly generated, \textit{mask application} faults happen when developers apply masks at the wrong computational stages (\eg{} post-softmax instead of pre-softmax), perform masking steps out of order, or combine multiple masks incorrectly. These issues can cause the model to assign attention scores to the wrong tokens, even when the mask tensor values look normal, creating detection challenges, especially when masks are updated dynamically. \textit{Dynamic mask mismatches} occur when systems fail to update or synchronize masks during operations like sequence-length changes or sliding window attention. Unlike static mask errors, these issues could lead to growing inconsistencies between how the mask should be updated and how it is actually updated during sequential operations. This is especially problematic in autoregressive models, where each step depends on the previous masking state. 

\indent\textbf{Root causes behind fault in QKV Projections \& Multi-head Handling.}
The parallel and distributed structure of QKV projections and multi-head attention creates several underlying root causes for faults in this category. \textit{Dimension mismatches} happen when implementations incorrectly split tensors across attention heads, creating incompatible shapes in internal computations. Dimension mismatches in QKV projections represent 10.3\% of all attention-specific root causes we identified, highlighting a significant architecture-specific vulnerability. Other root causes in this category include \textit{parameter initialization faults} and \textit{head interaction faults}. Improper weight initialization can disrupt head independence and cause gradient instability, while missing residual connections can create incorrect information flow between heads. These faults undermine the parallel processing capabilities that make multi-head attention effective, limiting head specialization and parameter optimization within the model architecture. In advanced adaptation scenarios, \textit{dynamic parameter registration faults} occur when newly added QKV projection or head weights are not registered for optimization, leaving these parameters excluded from training and freezing critical parts of the multi-head attention mechanism. 

\indent\textbf{Root causes behind fault in Attention Score Computation.}
We identify three main root causes of faults in attention score computation: implementation errors in the attention formula, mistakes in normalization steps, and incorrect handling of numerical precision. \textit{Implementation faults} in core attention formulas, especially missing key scaling factors like $\sqrt{d_k}$, can cause attention scores to be miscalibrated and improperly scaled. Our analysis indicates that errors in attention score computation (\eg{} using $\text{softmax}~(QK^T)$ instead of $\text{softmax}~(QK^T/\sqrt{d_k})$) can mimic overfitting, complicating the fault diagnosis process. \textit{Normalization faults} occur when softmax operations are applied incorrectly, skewing attention score distributions. A common error is applying dropout after, rather than before, normalization, which changes intermediate attention values and results in incorrect attention scores. We also found \textit{precision handling faults} from mixed-precision calculations account for 6.2\% of root causes of all attention-specific faults. These faults occur when numerical formats are not managed correctly, such as inconsistent upcasting or downcasting of tensors, or failing to address underflow and overflow of tensor values. These problems often start in low-level tensor operations or kernel design and can trigger numerical instability that propagates to later stages of the model. For example, in GitHub issue \#32570 for Mamba-2, incorrect handling of precision settings during training led to exploding gradients, which in turn disrupted model training~\cite{hf-issue32570}.

\indent\textbf{Root causes behind fault in Attention Kernel Selection \& Integration.}  
We noticed \textit{hardware incompatibility} and \textit{feature constraints} are common reasons for faults in specialized attention kernels, which are mostly used for optimization~\cite{dao2022flashattention}. These kernels often require unsupported device capabilities or reject valid input configurations due to kernel-specific limitations. More problematic are \textit{silent fallback faults}, where systems automatically downgrade to less efficient implementations without notification, which often leads to increased latency. These faults result from missing backend checks or misconfigured dispatch logic. For example, in FlashAttention~\cite{flashattention}, providing a non-null attention mask triggers a silent fallback to a slower backend, significantly increasing execution time~\cite{flashattention-issue1321}, exposing a critical blind spot in hardware-optimized ABNN implementations.

We also observed \textit{numerical precision faults}, such as improper quantization or incorrect rounding in kernel implementations, which can distort intermediate results depending on hardware-specific behavior. \textit{kernel memory management faults} include allocation or buffer overflows in optimized backends, especially when processing long sequences. These often stem from missing bound checks or improper indexing in memory management routines.

\indent\textbf{Root causes behind fault in Positional Encoding Integration.}
Attention’s critical dependence on the order of inputs creates vulnerabilities specific to model positional encoding. We found that \textit{indexing faults} often arise from incorrect or inconsistent position index calculations, which can lead to misaligned positional encoding during model execution. We also identified \textit{interpolation faults}, which occur when sequence lengths exceed those seen during training and the system applies invalid interpolation or extrapolates beyond supported positions. For example, using interpolation to handle sequences longer than 512 tokens in BERT, even though BERT was only trained on sequences up to 512~\cite{devlin2019bert}. We also observed \textit{relative position mismatches} caused by mistakes in position bucketing or offset calculations, which can assign incorrect relative positions to tokens during attention calculation.

\indent\textbf{Root causes behind fault in KV Cache Management.}
State management during inference introduces distinct technical challenges for attention mechanisms. We found \textit{cache invalidation} often comes from missing or improper updates to cached representations. \textit{Memory layout faults} result from incorrect allocation or indexing of cache structures, reducing storage efficiency. \textit{Update synchronization faults} occur when concurrent modifications to cache states lack proper coordination. These synchronization issues become more complex with \textit{cache-position mismatch}, which results from incorrect association between positional embeddings and cached content. We also observed \textit{distributed synchronization faults}, which happen when cache state consistency is not maintained across multiple devices. For example, in distributed inference, one GPU updates the KV cache while another still uses stale values, leading to mismatched outputs. We noticed that inadequate synchronization of concurrent KV cache updates in ABNNs sometimes leads to race conditions in autoregressive execution, increasing the risk of non-deterministic execution failures during sequence generation.

\indent\textbf{Root causes behind fault in Attention Variant Selection.}
Faults in attention variant selection are mainly caused by errors in implementation logic or incorrect parameter settings. We found that \textit{dynamic dispatch faults} occur when systems select suboptimal implementations for given inputs because of misconfigured dispatch logic. \textit{Variant configuration faults} happen when specialized attention types are set up with incorrect parameters, which can disable certain algorithmic features (\eg{} sliding-window processing). These configuration issues come from errors in code selection or parameter management within the attention mechanism. For example, in Github issue\#35335, a default setting turned off relative attention in DeBERTa, silently disabling the intended attention mechanism and making the model act like BERT~\cite{transformers-35335}.

\begin{rqboxgray}
\textbf{Summary in RQ2:} We identified 25 distinct root causes behind attention-specific faults, with mask generation errors (16.5\%) and QKV dimension mismatches (10.3\%) as the most frequent. Hardware constraints, precision handling errors, and KV cache synchronization errors also contribute notably, together accounting for the majority of highly severe failures in ABNNs.
\end{rqboxgray}

\subsection{Answering RQ3: Symptoms}
\label{sec:rq3_results}
RQ3 investigates how and where (in the DL pipeline) attention-specific faults are manifested. To answer RQ3, we captured and classified the symptoms of 292 faults by two dimensions: \textit{Observability} (\ie{} how easily the symptom is detected), \textit{Impact Surface} (\ie{} when or where a symptom appears) as follows.

\textbf{(a) Observability Patterns.} Following existing work~\cite{silentbugs-framework, aging-related-bugs, GRIST}, we categorize symptoms as \textit{explicit} (\ie{} immediate and observable errors, such as crashes, assertion failures, or runtime exceptions), \textit{silent} (\ie{} incorrect or unexpected outputs only detectable through an inspection)~\cite{silentbugs-framework, pytorchbugs}, and \textit{latent} (\ie{} hidden symptoms and only apparent after prolonged operation or in rare deployment conditions)~\cite{chen2021empirical,GRIST}. Our analysis shows that most attention-specific faults ($63.8$\%) demonstrate explicit symptoms, whereas $21.9$\% are silent, and the remaining $14.2$\% show latent symptoms. Despite this, we found that both silent and latent symptoms occur more often in ABNNs compared to our sample of classic DNNs.

\textbf{(b) Impact Surface: Runtime \& Computation.} Many symptoms are visible when an attention model is running, either during training or inference. Several issues, such as shape or dimension errors, produce immediate exceptions (\eg{} \texttt{RuntimeError: size mismatch for weight...}) and can be seen at a model's runtime. Explicit runtime symptoms also include \textit{numerical instability} (\ie{} NaN/Inf values in model output), and kernel compatibility errors
(\eg{} \texttt{RuntimeError: FlashAttention supports float16/bfloat16 only...}) and data type incompatibility errors. While analyzing attention faults, we also found other explicit runtime symptoms such as \textit{type propagation errors}, \textit{backend compilation failures}, \textit{autograd integration errors}, and \textit{memory access violations}, which occur as immediate crashes during training or inference (\eg{} \texttt{RuntimeError: CUDA error: an illegal memory access in...}).

Not all runtime symptoms are explicit; we identified 47 fault instances with silent manifestations during training or inference. One example is \textit{unexpected backend downgrade}, where the system silently falls back to a less optimized kernel, resulting in significant performance degradation (\eg{} increased latency) without any user-facing error or warning. For example, in issue\#12656, unsupported head size caused a silent fallback from \texttt{FlashAttention\-2} to the slower \texttt{XFormers} backend, increasing latency without any explicit warning~\cite{vllm-12656}.

Other silent runtime symptoms include \textit{causal mask ambiguity}, which manifests as improper weight distributions or tokens attending to unintended positions. These symptoms do not produce any explicit errors but can be observed by inspecting model outputs or attention maps. Another subtle symptom is \textit{frozen or partially updated model parameters}, which can be identified through weight inspection or abnormal training progress.

\textbf{(c) Impact Surface: Output Quality.} We find a range of silent degradation patterns in attention model outputs. These include \textit{attention divergence or collapse}, which appears as uniform or highly degenerate attention distributions~\cite{attentioncollapse, attentioncollapsecompress}, and \textit{attention distribution anomaly}, where the model's focus becomes overly concentrated on a few tokens~\cite{attentiondetails}. Both lead to less informative or coherent outputs, often visible in attention maps or generated sequences. We also distinguish between immediate symptoms, like repetitive outputs (“Hello Hello Hello”), and progressive output degradation, where output quality declines over longer sequences or extended inference. These issues may show up as abnormal repetition, loss of coherence, or a steady drop in performance metrics. Additional output-level symptoms include \textit{context bleeding}, where information from one sequence appears in another’s output, such as during batched inference without proper masking, and \textit{positional degradation}, where output quality varies abnormally with token position. Our comparative analysis shows that these output degradation symptoms (\eg{} attention collapse, context bleeding) are common in attention models but absent in our sample of classic DNN faults, highlighting their rarity outside attention architectures.

\textbf{(d) Impact Surface: Resource Management.} Resource management symptoms are observable effects of ABNN faults that disrupt the broader computational environment of the model. These include \textit{Out-of-Memory (OOM) errors} during attention computation, as well as problems that emerge in distributed or long-context scenarios, such as \textit{progressive output degradation} and \textit{non-deterministic generation}. Sometimes, these symptoms become apparent only after long sequences, with output quality dropping or results becoming inconsistent over time (\ie{} latent). Other resource management symptoms include \textit{silent performance regression}, where latency increases significantly without an error message, and \textit{post-quantization accuracy regression}, which manifests only as reduced output accuracy. In distributed settings, we observed \textit{node output mismatch} and \textit{output drift}, where identical inputs produce inconsistent or diverging results across compute nodes. For example:

\begin{lstlisting}[
    style=github,
    language=Python,
    caption={GPU Cache Issue (latent symptom)},
    captionpos=t,
    label={lst:gpu-cache}
]
# GPU0: Updated cache v3 & weights v2
# GPU1: Stale cache v2 & weights v1
output = distributed_attention(query, caches, params)
# Gradual degradation: ~10\% accuracy drop per 1K tokens
\end{lstlisting}

\begin{rqboxgray}
\textbf{Summary of RQ3:} Most attention-specific faults (63.8\%) demonstrate explicit symptoms, while 21.9\% are silent and 14.2\% show latent symptoms. We found that specific silent and latent symptom manifestations, such as attention collapse, context bleeding, and attention distribution anomalies, occur only in ABNNs and are not observed among symptoms caused by faults in classic DNNs.
\end{rqboxgray}

\subsection{Answering RQ4: Correlation}
\label{sec:rq4_results}
\begin{figure}[h]
\centering
\includegraphics[width=\columnwidth,alt={Sankey diagram mapping fault types, root causes, and symptoms}]{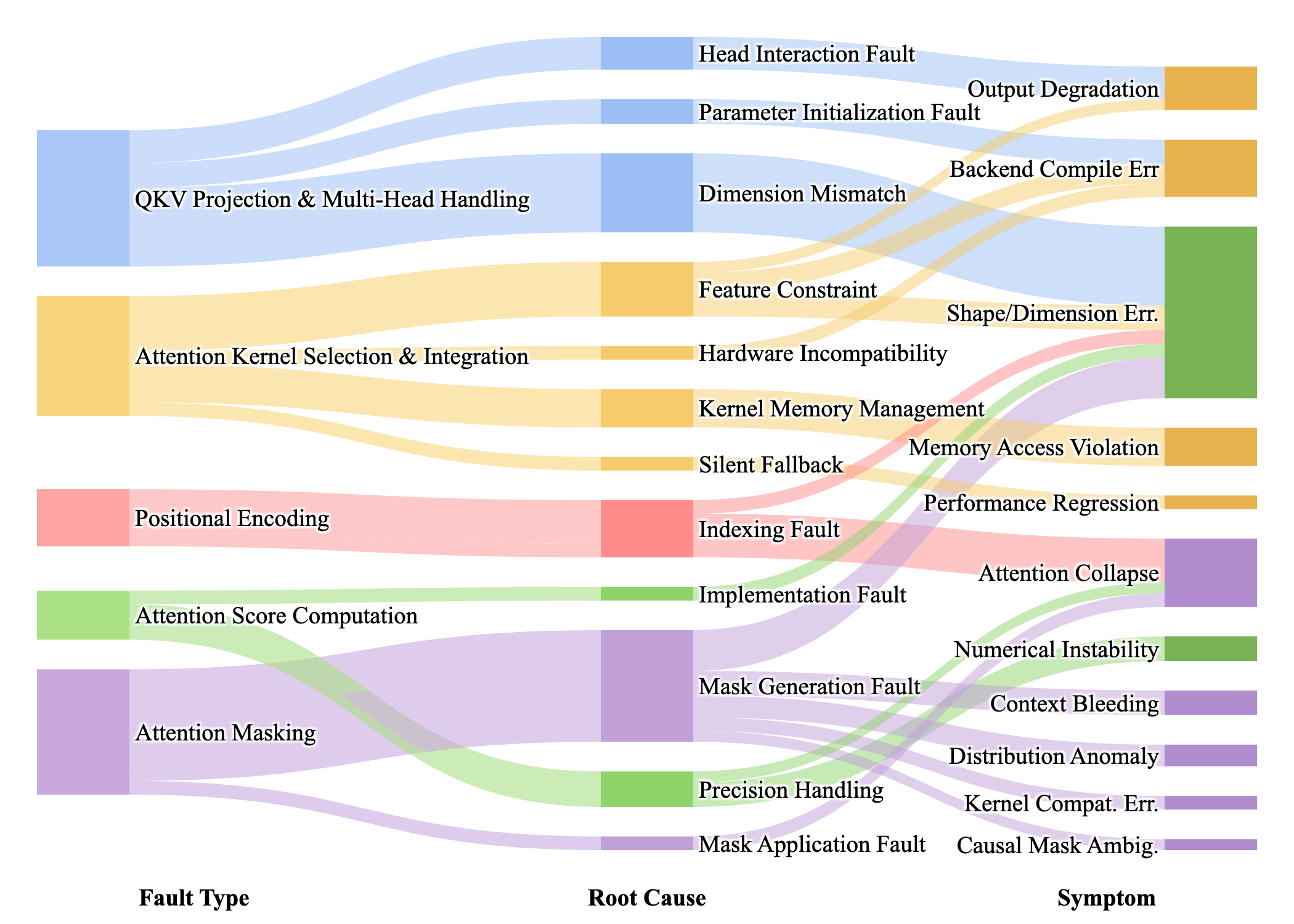}
\caption{Mapping among fault types, their corresponding root causes, and symptoms}
\label{fig:sankey}
\end{figure}

RQ4 investigates how symptoms and root causes are correlated in attention-specific faults. To answer RQ4, we analyzed the symptoms and root causes of 292 attention-specific faults, each recorded during our manual analysis (see~\ref{subsec:manual_analysis}). We examined symptom–root cause entries for each fault using co-occurrence analysis~\cite{dlbugcharacterstics, BugsDeepLearningCompilers} and identified 18 pairs, each supported by at least 12 faults (see Fig.~\ref{fig:sankey}). By inspecting the pair count distribution and testing different support thresholds, we selected 12 as the minimum value that filtered out rare, low-frequency pairs while retaining all recurring patterns. A chi-square test ($\chi^2 = 152.3$, $p < 10^{-20}$) confirmed a statistically significant association between symptom types and root cause categories.

Further analysis revealed that many of these pairs reflected broader, recurring patterns in attention faults. For example, multiple pairs link \textit{shape/dimension errors} as a symptom to underlying causes such as \textit{dimension mismatches}, \textit{cache-position mismatch}, and \textit{dynamic mask mismatches}. By aggregating related pairs, we derived four evidence-based diagnostic heuristics, each generalizing across multiple symptom–root cause pairings. Each heuristic is expressed using our taxonomy and takes the form, \textit{"if pattern X, then fault Y is likely”}, with both antecedent and consequent mapped to categories in our taxonomy. For each heuristic, we report support (the proportion of faults exhibiting the antecedent pattern) and confidence (the probability of the fault given the pattern), both calculated on our labeled dataset. Collectively, these four heuristics account for 33.0\% of attention-specific faults.

\paragraph{Heuristic 1:} This heuristic covers 10.0\% of attention-specific faults in our dataset. We observed that dimension mismatches in query ($Q$), key ($K$), or value ($V$) matrices caused attention failures, 67\% due to incompatible query-key dimensions, 24\% from incorrect multi-head dimension splitting, and 9\% from mismatched float types (\eg{} float16 vs. float32).
\begin{align}
\text{Attention(Q,K,V)} = \text{softmax}\left(\frac{QK^T}{\sqrt{d_k}}\right)V
\end{align}

These faults violate the requirements of the scaled dot-product attention mechanism (Equation 1), which demands $d_q = d_k$ for valid matrix multiplication and $d_m = n_h \times d_h$ for multi-head splitting.\\
\[
\textbf{Heuristic 1:}~
\left\{
\begin{array}{l}
\text{If}~d_q \neq d_k~\text{ or }~d_m \neq n_h d_h \\
\hspace{0.5em}\text{or }~\text{dtype}(Q) \neq \text{dtype}(K)~\text{or}~\text{dtype}(K) \neq \text{dtype}(V) \\
\hspace{2em}\text{or }~\exists\, x \in \{Q, K, V\} : x~\text{contains NaN/Inf} \\
\text{is strongly associated with fault in}\\
~\textcolor{red}{\textnormal{\textit{QKV Projection \& Multi-Head Handling.}}} \\
\end{array}
\right.
\]
\noindent
In our analysis, this pattern is most frequently associated with faults categorized as \textit{QKV Projection \& Multi-Head Handling} in our taxonomy.

\paragraph{Heuristic 2:} This heuristic explains 8.6\% of attention-specific faults. Errors in causal masking led to context leakage or attention collapse in autoregressive models such as GPT. Of these faults, 72\% resulted in uniform attention distributions (\ie{} low entropy, $H(W) < 0.1$), and 28\% involved incorrect position indexing in key-value caching. Causal attention relies on an upper-triangular mask to block future tokens, formally requiring $M_{i,j} = 0$ for $j > i$~\cite{vaswani2017attention}.
\[
\textbf{Heuristic 2:}~
\left\{
\begin{array}{l}
\text{If}~ H(W) < 0.1~\text{ or }~\sum_{i>j} M_{i,j} \neq 0~\\
\hspace{1em} \text{or }~\max(\text{indices}) \geq L\\
\hspace{1em} \text{ or }~\left|\sum_{j} W_{ij} - 1\right| > 10^{-6}~\forall i~\\
\text{is strongly associated with fault in}\\
~\textcolor{red}{\textit{Attention Masking.}}\\
\end{array}
\right.
\]
\noindent
Faults exhibiting this pattern are often classified as \textit{Attention Masking} in our taxonomy, based on the strong empirical association observed in our dataset.

\begin{table*}[h]
\centering
\caption{Support, confidence, and recall for each diagnostic heuristic, based on 292 annotated faults.}
\begin{tabular}{@{}lcccccc@{}}
\toprule
\textbf{Heuristic} & \textbf{Category} & \textbf{Total} & \textbf{Support (\#)} & \textbf{Support (\%)} & \textbf{Confidence (\%)} & \textbf{Recall (\%)} \\
\midrule
1 & QKV    & 64 & 27 & 9.2 & 93 & 42.2 \\
2 & Mask   & 73 & 25 & 8.6 & 91 & 34.2 \\
3 & Mask   & 73 & 20 & 6.8 & 87 & 27.4 \\
4 & Kernel & 54 & 15 & 5.1 & 90 & 27.8 \\
\bottomrule
\end{tabular}
\label{tab:heuristics}
\end{table*}

\paragraph{Heuristic 3:} This heuristic applies to 9.6\% of attention-specific faults. Here, attention mask errors were caused by shape mismatches (56\%), NaN or infinite values (31\%), or padding masks that allowed attention to invalid tokens (13\%). Masks control information flow via $\text{masked\_scores} = \text{scores} + \text{mask}$, requiring identical shapes for masks and attention scores and finite values to prevent numerical corruption. Padding masks must also align with the input sequence structure, so that all positions marked as padding ($P_j = 0$) are masked in every attention row. These checks prevent silent attention errors, such as models attending to padding tokens.
\[
\textbf{Heuristic 3:}~
\left\{
\begin{array}{l}
\text{If}~\text{shape}(M) \neq \text{shape}(S)~\\
\hspace{2em}\text{or}~\exists\, x \in M : x~\text{is NaN/Inf}~\\
\hspace{2em}\text{or}~\exists\, (i, j)~\text{s.t.}~P_j = 0~\text{ and }~M_{i,j} \neq \text{mask\_value}~\\
\text{is strongly associated with fault in}\\
~\textcolor{red}{\textit{Attention Masking.}}\\
\end{array}
\right.
\]
\noindent
We found this pattern most frequently in faults categorized as \textit{Attention Masking} from our taxonomy.

\paragraph{Heuristic 4:} This heuristic addresses 4.8\% of attention-specific faults. Optimized attention kernels (\eg{} FlashAttention) failed due to sequence lengths exceeding kernel limits (71\%) or batch sizes causing memory overflows (29\%). These kernels, with quadratic complexity $O(BL^2d_h)$, pre-allocate memory based on the maximum supported sequence length ($L_{\text{max}}$) and available GPU memory ($M_{\text{avail}}$)~\cite{memoryefficientflashattn}. We enforce checks to ensure sequence length ($L$) does not exceed $L_{\text{max}}$, and that memory usage ($B \times L^2 \times d_h$) stays within available hardware capacity. These conditions prevent CUDA errors or segmentation faults in long-sequence scenarios (\eg{} $L = 16\text{k}$ exceeding $L_{\text{max}} = 8\text{k}$).
\[
\textbf{Heuristic 4:}~
\left\{
\begin{array}{l}
\text{If}~L > L_{\text{max}}~\\
\hspace{2em}\text{or}~B \times L^2 \times d_h > M_{\text{avail}}~\\
\text{is strongly associated with fault in}\\
~\textcolor{red}{\textit{Attention Kernel Selection \& Integration.}}\\
\end{array}
\right.
\]
\noindent
This pattern is highly indicative of faults in the~\textit{Attention Kernel Selection \& Integration} category.

Table~\ref{tab:heuristics} reports the support, confidence, and recall of each heuristic. Confidence remains high (87–93\%), reflecting strong precision, while recall ranges from 27–42\% due to the long-tail nature of ABNN faults~\cite{chen2023toward}, where few patterns recur often, but many are rare and diverse. Our heuristics prioritize reliability over exhaustive coverage, capturing the most recurrent patterns with high confidence. They offer a practical foundation for taxonomy-driven fault diagnosis in ABNN.

\begin{rqboxgray}
\textbf{Summary of RQ4:} We identified four recurring patterns that generalize 18 common symptom–root cause pairs. Expressed as diagnostic heuristics, they explain 33.0\% of all attention-specific faults with high confidence (87–93\%) and moderate recall across major fault categories like \textit{QKV Projection}, \textit{Attention Masking}, and \textit{Attention Kernel Selection \& Integration}. 
\end{rqboxgray}

\section{Threats To Validity}
Threats to \emph{construct validity} concern whether our fault classification accurately reflects real-world ABNN faults~\cite{smith2005construct}. We mitigated this by applying established guidelines~\cite{miles1994qualitative}, having two authors independently label all faults, and resolving any disagreements through discussion. Threats to \emph{internal validity}~\cite{internalvalidity} stem from possible bias or experimental errors. Our high inter-rater agreement might have mitigated such threats. Threats to \emph{external validity} relate to the generalizability of our findings~\cite{findley2021external}. While our dataset spans ten frameworks, findings may not extend to all ABNN architectures or future models. 

\section{Related Work}
Early taxonomies~\cite{faulttaxonomy,dlbugcharacterstics} classified training faults, API misuse, and tensor errors in classic DNN architectures, establishing a foundation for deep learning faults. Later studies~\cite{comprehensiveFrameworkBug, faultsinLibrary, performance-bugs-framework,aging-related-bugs,faultsinCompiler, MLoptimizationBug} explored specific parts of the DL pipeline. Research on deep learning frameworks analyzed component-level bugs in systems like TensorFlow and PyTorch~\cite{comprehensiveFrameworkBug,faultsinLibrary}. Other studies focused on performance and memory issues~\cite{performance-bugs-framework,aging-related-bugs}, compiler and deployment faults~\cite{comilerbugsempirical,faultsinCompiler,chen2021empirical,dependencyBugs}, user-code and API misuse~\cite{pytorchbugs,silentbugs-framework,morovati2024bug}, and optimization errors~\cite{MLoptimizationBug,morovati2023bugs}. These works have advanced fault analysis but did not address the unique challenges of attention mechanisms. Table~\ref{tab:attention_coverage} shows that prior studies either completely overlook attention-specific faults or might only partially cover them. \citet{performance-bugs-framework} focus on performance issues and may include attention-related inefficiencies such as slow kernels. Similarly,~~\citet{pytorchbugs} analyze silent bugs and flag missing softmax operations in attention modules. However, these works do not investigate attention-specific root causes (\eg{} mask misuse, QKV misalignment), which are essential for fault diagnosis. To the best of our knowledge, no prior taxonomy systematically addresses faults unique to ABNNs. We address this gap by analyzing 555 real-world ABNN faults, identifying seven novel attention-specific fault types and 25 root causes (see Fig.~\ref {fig:taxonomy}), and mapping root causes to observable symptoms to derive actionable rules for fault diagnosis.

\begin{table}[H]
\centering
\caption{Scope of Prior Fault Studies \& Coverage of Attention-specific Faults}
\resizebox{0.9\columnwidth}{!}{%
\begin{tabular}{@{}p{0.6cm}p{6.8cm}p{1.2cm}@{}}
\toprule
\textbf{Study} & \textbf{Primary Fault Scope} & \textbf{Coverage} \\
\midrule
\rowcolor{gray!15}
\multicolumn{3}{c}{\textit{Framework Bugs}} \\
\cite{aging-related-bugs} & Aging/resource leaks in core DL frameworks & No \\
\cite{comprehensiveFrameworkBug} & API, logic, data, numeric bugs in framework internals across 8 DL libraries & No\\
\cite{silentbugs-framework} & Silent weight/gradient issues in Keras \& TF & No \\
\cite{performance-bugs-framework} & Performance regressions (memory/kernel inefficiencies) in DL frameworks (TF, PyTorch) & Partial\\
\cite{faulttaxonomy} & General DNN library misuse and runtime issues & No \\
\midrule
\rowcolor{gray!15}
\multicolumn{3}{c}{\textit{Compiler \& Deployment}} \\
\cite{comilerbugsempirical} & Environment and memory bugs in DL compilers & No \\
\cite{faultsinCompiler} & Shape \& optimization faults in DL compilers & No \\
\cite{chen2021empirical} & Deployment and integration faults in DL apps & No \\
\cite{dependencyBugs} & Version \& driver configuration faults in DL stack & No \\

\midrule
\rowcolor{gray!15}
\multicolumn{3}{c}{\textit{User Code \& APIs}} \\
\cite{pytorchbugs} & Silent user-code faults in PyTorch (operation omissions/misplacement/non-convergence) & Partial\\
\cite{morovati2024bug} & API misuse \& tensor alignment faults in ML apps & No \\
\cite{dlbugcharacterstics} & Misuse of DL libraries, data and logic faults & No \\
\cite{faultsinLibrary} & Symptom/cause taxonomy for TF library faults & No \\

\midrule
\rowcolor{gray!15}
\multicolumn{3}{c}{\textit{Performance \& Optimization}} \\
\cite{understandingPerformanceBugsinDL} & Performance anti-patterns, slow training, OOM & No \\
\cite{MLoptimizationBug} & Model optimization faults & Partial\\
\cite{morovati2023bugs} & Benchmark, ML component bugs (TF/Keras) & No \\

\midrule
\rowcolor{gray!15}
\multicolumn{3}{c}{\textit{Other}} \\
\cite{faultsinRL} & Faults in deep-reinforcement learning modules & No \\
\bottomrule
\end{tabular}
}
\label{tab:attention_coverage}
\end{table}

\section{Conclusion \& Future Work}
In this paper, we present the first comprehensive empirical study of faults in attention-based neural networks, based on 555 real-world issues from ten frameworks. Our taxonomy reveals seven attention-specific fault categories and 25 root causes, with over half of ABNN faults stemming from mechanisms unique to attention models. We find that silent and latent faults are significantly more frequent in ABNNs than in classic DNNs. Building on the correlation between symptoms and root causes, we propose four diagnostic heuristics that collectively help identify 33\% of all attention-specific faults. In future work, we plan to develop a comprehensive framework for diagnosing attention-related faults.
\balance
\bibliographystyle{ACM-Reference-Format}
\bibliography{ref}
\end{document}